\documentclass[conference]{IEEEtran}
\IEEEoverridecommandlockouts
\usepackage{cite}
\usepackage{amsmath,amssymb,amsfonts}
\usepackage{algorithmic}
\usepackage{graphicx}
\usepackage{textcomp}
\usepackage{xcolor}
\usepackage{float}
\usepackage{booktabs}
\usepackage{caption}
\usepackage{amsmath}
\usepackage{epstopdf}
\usepackage{epsfig}
\usepackage[ruled]{algorithm2e}
\usepackage{multirow}
\usepackage{xcolor}
\usepackage{colortbl} 
\usepackage{hhline} 
\usepackage{longtable}
\usepackage{lscape}
\usepackage{float}
\usepackage[colorlinks=true, linkcolor=blue, citecolor=blue, urlcolor=blue]{hyperref}
\usepackage[table,xcdraw]{xcolor}
\usepackage{multirow}
\usepackage{tabularx} 
\usepackage{enumerate}
\usepackage{cleveref}
\usepackage{stfloats}
\usepackage{graphicx}
\usepackage{subfig}
\usepackage{graphicx}
\usepackage{titlesec}
\usepackage{indentfirst}
\usepackage{caption}
\def\BibTeX{{\rm B\kern-.05em{\sc i\kern-.025em b}\kern-.08em
		T\kern-.1667em\lower.7ex\hbox{E}\kern-.125emX}}
\begin{document}
	
	\title{Enhancing Low-Altitude Airspace Security: MLLM-Enabled UAV Intent Recognition}
	
	\author{Guangyu Lei, Tianhao Liang,~\IEEEmembership{Member,~IEEE}, Yuqi Ping, Xinglin Chen, Longyu Zhou,~\IEEEmembership{Member,~IEEE},
\\Junwei Wu, Xiyuan Zhang, Huahao Ding, Xingjian Zhang,~\IEEEmembership{Member,~IEEE}, 
Weijie Yuan, ~\IEEEmembership{Senior Member,~IEEE},\\Tingting Zhang,~\IEEEmembership{Member,~IEEE} and Qinyu Zhang,~\IEEEmembership{Senior Member,~IEEE}
		\thanks{Guangyu Lei, Tianhao Liang, Yuqi Ping, Xinglin Chen, Junwei Wu, Xiyuan Zhang, Huahao Ding, Xingjian Zhang, Tingting Zhang and Qinyu Zhang are with the School of Information Science and Technology, Harbin Institute of Technology (Shenzhen), Shenzhen 518055, China, Xingjian Zhang,  Tingting Zhang and Qinyu Zhang are also with the Guangdong Provincial Key Laboratory of Space-Aerial Networking and Intelligent Sensing, Shenzhen, China (e-mail: GuangyuLei@stu.hit.edu.cn, liangth@hit.edu.cn, pingyq@stu.hit.edu.cn, chenxinglin@stu.hit.edu.cn, 220210419@stu.hit.edu.cn, 220210924@stu.hit.edu.cn, hitszdhh@163.com, x.zhang@hit.edu.cn, zhangtt@hit.edu.cn,  zqy@hit.edu.cn). Xingjian Zhang, Tingting Zhang and Qinyu Zhang is also with PengCheng Laboratory, Shenzhen 518055, China. Longyu Zhou is with Information Systems Technology and Design, Singapore University of Technology and Design (e-mail: zhoulyfuture@outlook.com). Weijie Yuan is with the School of System Design and Intelligent Manufacturing, Southern University of Science and Technology, Shenzhen, China (e-mail: yuanwj@sustech.edu.cn).
		}
	}

	\maketitle
	
	\begin{abstract}
		
	The rapid development of the low-altitude economy emphasizes the critical need for effective perception and intent recognition of non-cooperative unmanned aerial vehicles (UAVs). The advanced generative reasoning capabilities of multimodal large language models (MLLMs) present a promising approach in such tasks. In this paper, we focus on the combination of UAV intent recognition and the MLLMs. Specifically, we first present an MLLM-enabled UAV intent recognition architecture, where the multimodal perception system is utilized to obtain real-time payload and motion information of UAVs, generating structured input information, and MLLM outputs intent recognition results by incorporating environmental information, prior knowledge, and tactical preferences. Subsequently, we review the related work and demonstrate their progress within the proposed architecture. Then, a use case for low-altitude confrontation is conducted to demonstrate the feasibility of our architecture and offer valuable insights for practical system design. Finally, the future challenges are discussed, followed by corresponding strategic recommendations for further applications.
	\end{abstract}
	
	\begin{IEEEkeywords}
		UAV, multimodal fusion perception, multimodal large language models, intent recognition.
	\end{IEEEkeywords}

	\section{Introduction}
\IEEEPARstart{W}{ith} the increasing development of unmanned aerial vehicle (UAV) technology, new opportunities have opened up in the low-altitude airspace. UAVs significantly enhance communications and sensing capabilities and enable critical applications such as aerial logistics delivery and environmental monitoring \cite{LiangT,UAV2, LT1, LT2, LT3, LT4}. However, the impacts of UAV proliferation are inherently dual-edged. While UAVs are widely utilized for civilian and industrial applications, non-cooperative UAVs may also exhibit behaviors such as trajectory conflicts or unauthorized reconnaissance, posing significant risks to low-altitude safety. Therefore, developing effective UAV intent recognition systems is essential to safeguard the safety and reliability of airspace usage.

Compared with manned aircraft, UAVs are generally smaller in size, slower in speed, and operate at lower altitudes, thus falling into the category of {\it\textbf{ low, slow, and small (LSS)}} targets. The inherent characteristics of LSS targets introduce substantial challenges for effective detection and tracking. Existing investigations have explored various sensing modalities for UAV surveillance, including visible RGB cameras, radar, infrared (IR), acoustic, and radio-frequency (RF) signals. Nevertheless, each single-modality approach exhibits limitations. RF-based methods can detect UAV presence but provide limited kinematic information. Acoustic-based methods have short effective ranges and are vulnerable to environmental noise. Vision-based methods suffer from performance degradation under poor lighting or adverse weather conditions. As such, reliance on a single sensor type often results in insufficient robustness in complex environments. This has driven a growing interest in multimodal fusion approaches.

Beyond perception tasks such as detection and tracking, the inference of UAV intent is equally crucial for the operation of the monitoring system. By analyzing UAV trajectories, formation patterns, and operational contexts, it is possible to infer their mission objectives and potential intentions, thereby enabling the formulation of timely countermeasures. Some existing works have attempted to extract UAV intent based on predefined labels and task flows, using supervised classification models trained on limited datasets \cite{GAI}. However, such methods suffer from the poor scalability to unstructured or adversarial scenarios, and often lack the flexibility to capture complex intent patterns accurately. Moreover, most existing solutions require fixed-format, complete input data, limiting their operational speed and adaptability in time-critical situations.

Multimodal large language models (MLLMs) offer significant advantages in addressing these challenges\cite{MLLM2,MLLMreview}. By effectively integrating heterogeneous data, including optical imagery, IR thermal maps, radar returns, and RF spectrum signatures, MLLMs can achieve cross-modal semantic alignment and complementary information enhancement, yielding more comprehensive scene information. Leveraging knowledge acquired during large-scale pretraining, MLLMs can perform complex reasoning and hierarchical analysis for intent recognition tasks with only minimal prompt-based priors, thereby reducing the cost of task adaptation and model retraining. Because of flexible prompt design, MLLMs can dynamically adapt to different mission objectives and tactical preferences, enhancing adaptability and scalability in diverse operational and environmental contexts. These capabilities position MLLMs as a promising foundation for building intelligent, generalizable, and reasoning-capable UAV intent recognition systems.

Motivated by the above issues, we explore the integration of multimodal information fusion and MLLMs-enabled UAV intent recognition. First, we present an MLLM-enabled UAV intent recognition system, comprising a multimodal fusion UAV perception system and an intent recognition module. Specifically, it introduces the principles of various sensors used in forming the perception system, outlining their advantages and limitations. Subsequently, we present the basic architecture and core functionalities of MLLMs, emphasizing how they can adaptively perform UAV intent recognition in complex environments. Moreover, we review the related works in UAV perception and intent recognition and present a use case for low-altitude confrontation, in which self UAVs collect real-time perception information on enemy UAVs and generate structured input to be uploaded to a cloud-based MLLM, producing intent recognition results and recommendations by incorporating contextual knowledge, prior information, and tactical preferences. Finally, future challenges are discussed based on the proposed architecture, followed by strategic recommendations.

%
	
	\section{Architecture of MLLM-Enabled UAV Intent Recognition System}
\begin{figure*}[htbp]
	\centering
	\includegraphics[scale=0.45]{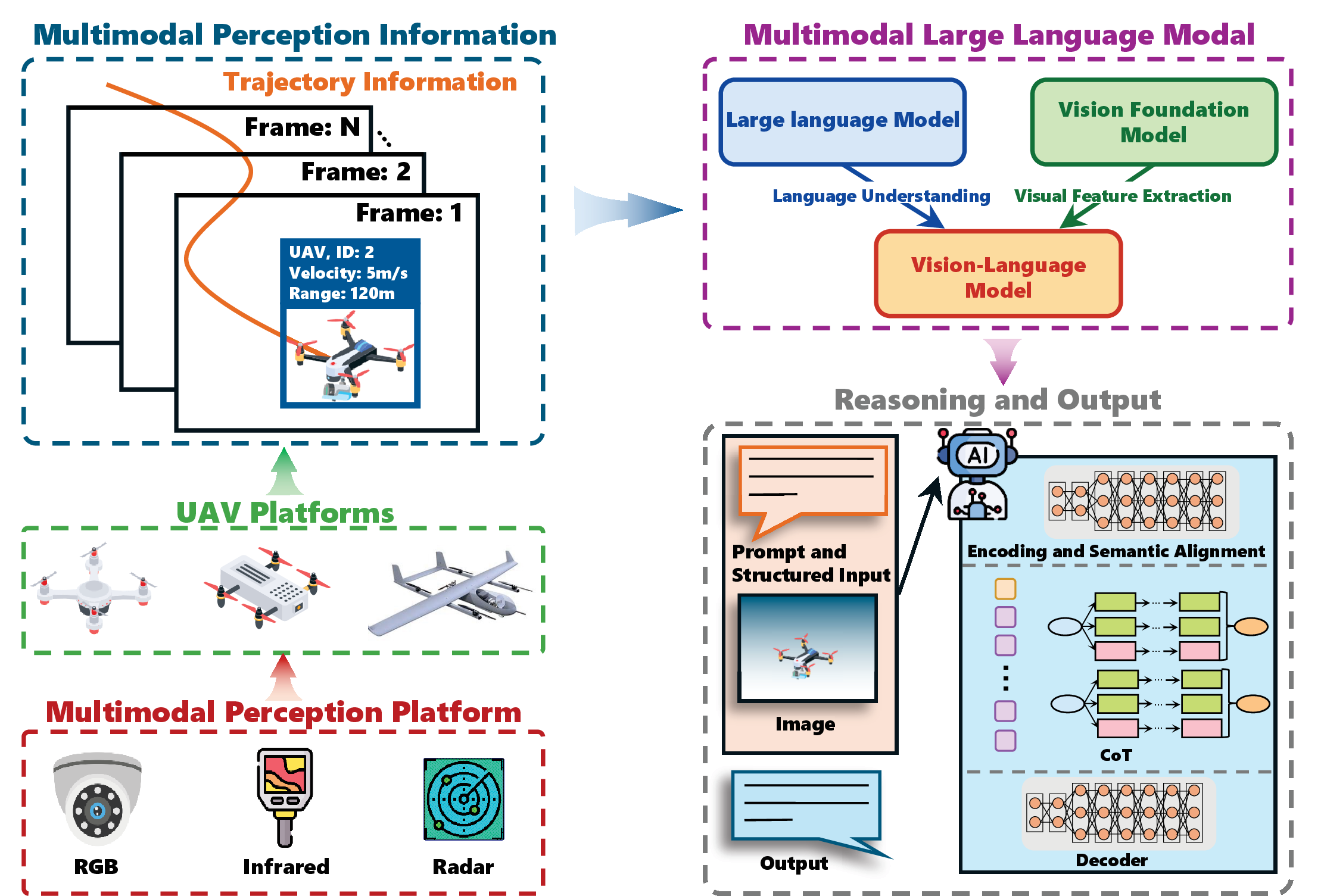} 
	\caption{The MLLM-enabled UAV intent recognition system architecture.}
	\label{Fig_1}
\vspace{-20pt} 
\end{figure*}

	The proposed architecture of the MLLM-enabled UAV intent recognition system is illustrated in Fig. \ref{Fig_1}, comprising a multimodal UAV perception module and an MLLM-enabled inference module. The perception module is responsible for acquiring comprehensive UAV information in real time as structured input. In contrast, the inference module leverages MLLM to perform reasoning and outputs the potential actions and intentions of enemy UAVs. A detailed introduction to each module will be provided as follows\footnote{In this section, several sensors that are currently widely used, possess strong anti-interference capabilities, and can provide relatively accurate perception information over long distances are presented.}.
	\subsection{UAV Perception}
	\subsubsection{RGB Camera} 

RGB cameras capture color image data as standard visual sensors. In UAV detection and countermeasure tasks, deep learning has become the mainstream for image processing, enabling neural networks to extract complex features adaptively. Visual detection offers advantages in speed, efficiency, range, and accuracy. The YOLO series \cite{YOLO} is a classic algorithm capable of regressing target positions and classes at hundreds of frames per second, while emerging methods can further improve detection accuracy. However, visual detection is limited by lighting conditions such as foggy weather or nighttime, becoming a key factor affecting detection effectiveness.

	\subsubsection{Infrared}

IR sensors capture mid to long wave radiation differences, complementing RGB in complex weather, reducing reliance on visible light, and maintaining high contrast in low-visibility conditions. Like RGB cameras, they output two-dimensional (2D) images and can employ similar CNNs, with public RGB-IR datasets supporting fusion research \cite{ANTIUAV}. However, IR performance is reduced in water vapor absorption, such as heavy rain or fog. Moreover, Limited texture in thermal images makes it hard to identify UAV types or payloads, and weak thermal signatures in small UAVs with efficient cooling further degrade detection.

	\subsubsection{Radar}

Airborne radars for UAVs include Light Detection and Ranging (LiDAR) and millimeter-wave (mmWave) radar. LiDAR provides centimeter-level 3D accuracy and dense point clouds, but suffers from scattering and absorption in fog or rain. mmWave radar, with intense penetration and all-weather capability, can exploit rotor-induced micro-Doppler for UAV discrimination and benefit from public datasets for CNN or temporal network. The maximum unambiguous radar detection range depends on the transmit power. For small UAV platforms, limited onboard power, battery capacity, and payload restrictions, airborne radars usually have constrained detection range and a small number of antenna elements, resulting in limited angle measurement capability for mmWave radar.

\subsubsection{Multimodal Fusion}

Multimodal fusion integrates data from heterogeneous sensors for robust UAV detection, identification, and tracking, leveraging different modalities' complementary strengths. By coordinating modalities, weather sensing, and motion measurement, radar can compensate for the shortcomings of visual sensors in low-visibility scenes. In contrast, high-resolution visual data can improve angle estimation and classification accuracy for distant targets, thus extending the detection range of system. Motion information derived by radar, specifically range and velocity, is fed back to the fusion center, providing prior motion constraints for vision-based detectors using convolutional or temporal networks. Continuous radar motion measurements support multi-target tracking algorithms such as Kalman filtering, probabilistic data association, and factor graph optimization, enabling continuous trajectory estimation and reliable intent recognition input.

	\subsection{MLLM-Enabled Intent Recognition}

\subsubsection{Model Architecture and Semantic Alignment}

A MLLM typically consists of a Vision Foundation Model (VFM), a Vision-Language Model (VLM), and a Large Language Model (LLM). The VFM is responsible for feature extraction from this architecture's perceptual modalities, such as images and videos. At the same time, the LLM serves as the core for reasoning and control, leveraging its strong language understanding and generation capabilities. MLLMs align different modalities into a unified semantic space via linear projection, cross-attention, or Q-Former modules. This semantic alignment enables the model to understand and integrate heterogeneous inputs jointly. Compared with traditional single-modality models, MLLMs offer greater environmental adaptability and task generalization, making them well-suited for multi-source perception in complex scenarios.

\subsubsection{Reasoning Ability and Task Generation}

The core strength of MLLMs lies in their ability to generate complex semantic reasoning and tasks, which the integrated LLM primarily drives. The LLM can process language inputs and reason over multimodal content through a Chain-of-Thought (CoT) mechanism, producing logical and coherent outputs. In UAV intent recognition scenarios, MLLMs can jointly model multimodal inputs, such as images, trajectories, and mission texts, to identify threats, infer target intent, and generate appropriate task responses or scheduling suggestions. Compared with traditional single-modal or rule-based systems, this reasoning process is more interpretable, semantically complete, and execution-flexible, supporting a full-loop {\it \textbf{perception–understanding–reasoning–output}} pipeline for intelligent task handling.

\subsubsection{Prompt Guidance and Context Control}

Prompts serve as the critical bridge between human intent and the model’s internal understanding, functioning as the core input method for enabling diverse MLLM tasks. Well-designed prompts can specify task format, output structure, semantic style, and reasoning strategy. In practice, users can input scenario descriptions, task settings, tactical preferences, and engagement rules as context to steer the reasoning direction and outputs of model. Prompt design determines the model’s focus and response strategy, and can also constrain outputs to structured formats for easier downstream fusion and execution.

\section{Related Works}

This section reviews research on UAV perception based on multimodal fusion and existing studies on UAV intent recognition using data from multiple modalities.

\begin{table*}[!t]
	\centering
	\includegraphics[scale=0.2]{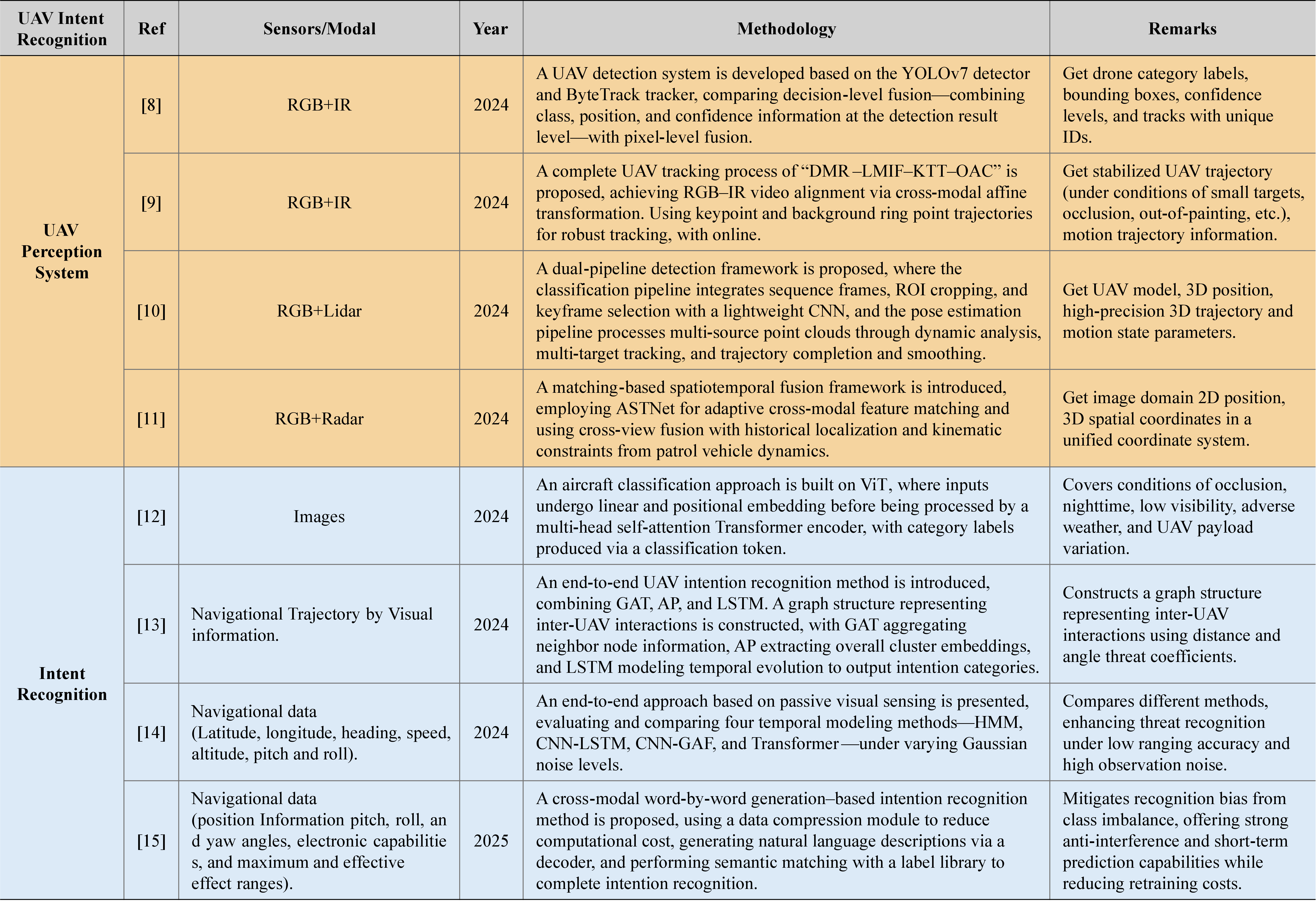} 
	\caption{The summarization of related works on multimodal UAV perception and intent recognition.}
	\label{tab_1}
\vspace{-10pt}

\end{table*}

\subsection{Multimodal-Driven UAV Perception}
In multimodal UAV perception research, various solutions integrating RGB camera, IR, radar, and positioning data have been proposed to improve detection accuracy and tracking stability of small UAVs across diverse scenarios and operational states.

The UAV detection system based on the YOLOv7 detector and ByteTrack tracker is proposed in Ref. \cite{IRRGB1}, utilizing RGB and IR dual-modal data. Two fusion strategies are compared, including decision-level fusion that combines category, location, and confidence at the detection output level, and pixel-level fusion that performs image registration and integration across modalities. Both approaches significantly enhance detection and tracking performance across scenarios, outputting UAV class labels, bounding boxes, confidence scores, and unique-ID trajectories. A dual-pipeline detection framework is presented in Ref. \cite{RGBRADAR1}, integrating visual and point cloud data. The classification pipeline processes image sequences, Region of Interest (ROI) cropping, and keyframe selection through a lightweight CNN for model identification. The pose estimation pipeline leverages multi-source point clouds from binocular fisheye cameras, conical 3D LiDAR, omnidirectional 3D LiDAR, and millimeter-wave radar. Dynamic point cloud analysis, multi-target tracking, and trajectory completion yield high-precision 3D trajectories and motion parameters, ultimately outputting UAV type, 3D position, and the whole trajectory. A matching-based spatiotemporal fusion framework for swarm intent recognition is proposed in Ref. \cite{RGBRADAR2}. The Attention-based Spatiotemporal Network (ASTNet) adaptively matches cross-modal features from radar and RGB cameras. At the same time, cross-view fusion integrates historical localization and kinematic constraints from patrol vehicles’ dynamic viewpoints to recover occluded targets. A complete counter-UAV tracking pipeline is proposed in Ref. \cite{IRRGB2}, comprising Dynamic Multimodal Registration (DMR), Layered Multimodal Information Fusion (LMIF), Keypoint Trajectory Tracking (KTT), and Online Adaptive Calibration (OAC). Cross-modal affine transformation aligns RGB and IR videos at the pixel level, while layered fusion of base and detail components enhances target saliency. Keypoint and background ring point motion replace appearance templates for robust tracking, and online correction is applied when point sparsity or geometric degradation occurs. This enables stable detection and tracking under small target size, occlusion, and out-of-view conditions.  A combination of phased-array pulse-Doppler radar and RGB cameras provides pixel-level location with confidence scores and 3D spatial coordinates in a unified reference frame, enabling motion range prediction under long-term occlusion.

A summary of related work is provided in Table \ref{tab_1}. Overall, these works leverage the complementary strengths of multimodal sensors and multi-level feature fusion strategies to achieve high-precision detection, robust tracking, and full-space localization of UAVs, from single targets to swarms. They provide a solid perception foundation and data support for MLLM-based UAV intent recognition in complex airspace environments.

\subsection{UAV Intent Recognition}

In recent years, research on UAV intent recognition and intent recognition in complex airspace has advanced significantly in feature modeling, temporal relation learning, and robustness enhancement.

A Vision Transformer (ViT)-based aircraft classifier is proposed in Ref. \cite{VIT}, segmenting input images into fixed-size patches for multi head self attention encoding, with a classification token outputting the label. On a five-class custom dataset, including malicious UAV, civilian UAV, airplane, helicopter and birds, under challenging visibility, occlusion, and payload conditions, it achieved high accuracy and strong robustness. Addressing the limitation of single-UAV, instantaneous feature-based methods is proposed in Ref. \cite{GAF} by proposing an end-to-end swarm intent recognition framework combining a Graph Attention Network (GAT), Attention Pooling (AP), and Long Short-Term Memory (LSTM). Interaction graphs are built using distance and angle threat coefficients, aggregated via GAT, globally embedded with AP, and temporally modeled with LSTM.  A passive RGB-based end-to-end intrusion recognition and interception method is proposed in Ref. \cite{GNN}, deriving trajectories via category-based ranging and the pinhole camera model. In Gazebo simulations with cruising, loitering, and evasion behaviors, CNN-GAF and Transformer models outperformed Hidden Markov Model (HMM) and CNN-LSTM in high-noise settings. Addressing long-sequence UAV behavior classification under class imbalance with a cross-modal token-by-token generation approach is proposed in Ref. \cite{Generation}. The compression module first reduces computation, then the Transformer encodes seven behavioral features, and finally the decoder generates natural language descriptions matched to a label library. Temporal smoothing, unsupervised contrastive learning, and cross-modal matching pre-training are provided to improve feature extraction and alignment. Tests on an imbalanced ten-intent wargaming dataset showed improved robustness, short-term prediction, and adaptability without retraining for label set expansion.

Considering all the factors, the existing research provides several implementations for UAV intent recognition on various modalities, offering scalable and adaptable solutions for MLLM-enabled UAV intent recognition in complex operational environments.

\section{Use Case: Low-Altitude Confrontation}

\begin{figure*}[!t]
	\centering
	\includegraphics[scale=0.18]{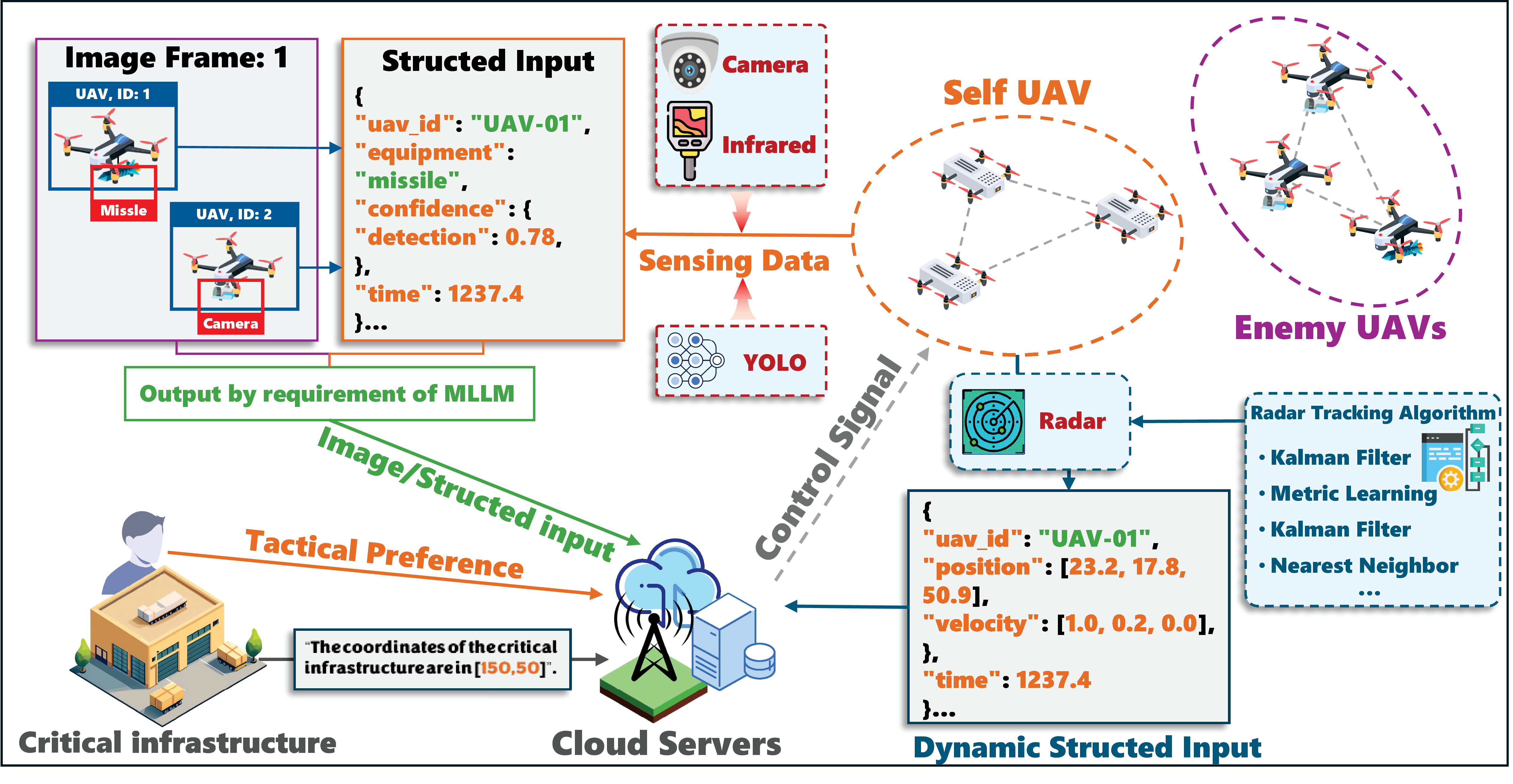} 
	\caption{The system architecture of the use case. Important prior information can be provided as prompts to the MLLM. The self UAV obtains multimodal perception results of enemy UAVs in the air and transmits them to the MLLM. The MLLM can then recognize the intent of the enemy UAVs and provide potential recommendations.}
	\label{Fig_2}
\vspace{-20pt} 
\end{figure*}

In this use case, the proposed architecture is utilized to enable self UAVs to extract payload and motion information of adversarial UAVs via multimodal perception devices and transmit it to the MLLM for intent recognition. Fig. \ref{Fig_2} illustrates the overall environment and architecture of the case, self UAVs acquire data on adversarial UAVs through multimodal sensors and transmit it via communication links to the cloud, where intent recognition tasks are performed. We will show case on how MLLM integration leverages multimodal perception results to overcome the limitations of traditional UAV intent recognition methods.

\begin{figure*}[!t]
	\centering
	\includegraphics[scale=0.11]{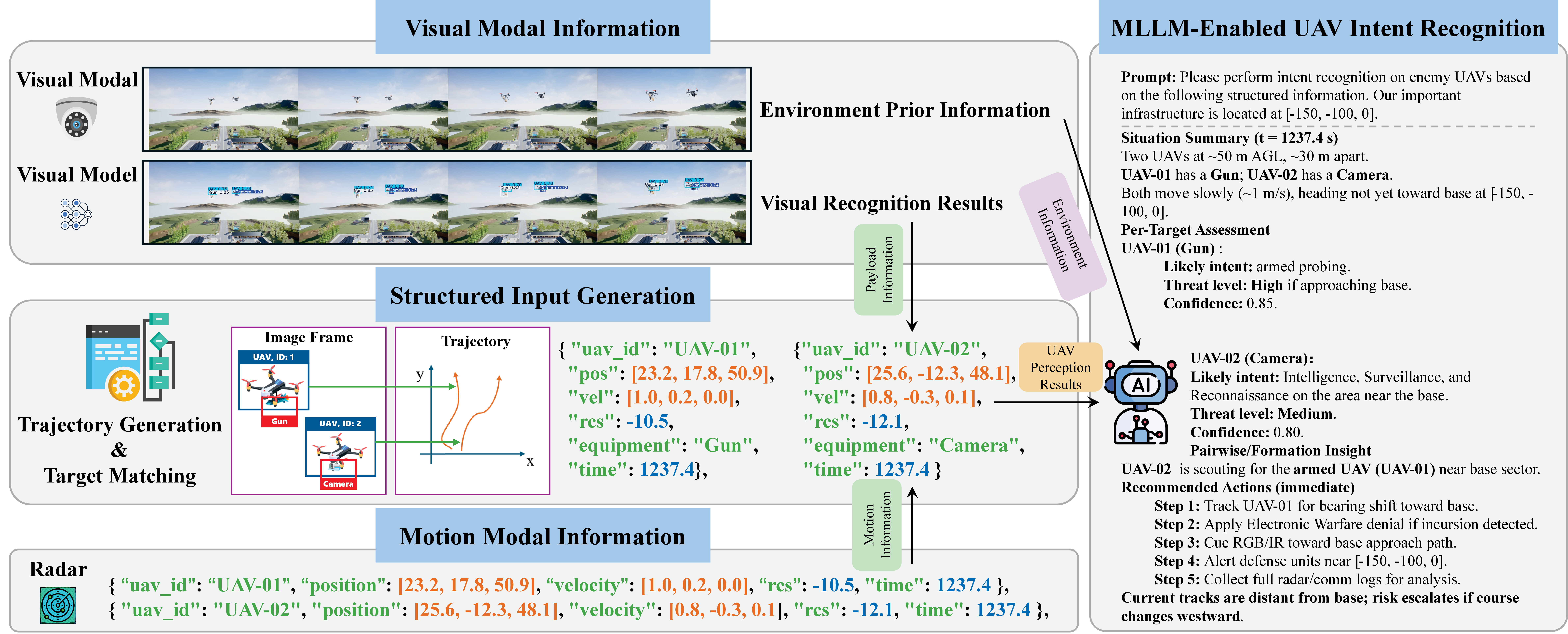} 
	\caption{The demonstration of use case, by inputting visual recognition results, motion information, scene images (the first-frame RGB camera image fed into the MLLM), and prior information, the MLLM can infer the potential intent of each enemy UAV and provide recommendations for the next course of action.}
	\label{Fig_3}
\vspace{-20pt} 
\end{figure*}

The intent recognition scenario is built on the open-source UAV simulation platform AirSim, with the Qwen-vl-plus MLLM deployed in the cloud. Self UAVs are equipped with RGB cameras and radar to detect multiple adversarial UAVs carrying weapons or reconnaissance equipment and approaching the defended area\footnote{As the lighting conditions configured in the AirSim platform are relatively ideal, only the RGB camera is used for the demonstration of the visual modality.}.  In the experimental setup, radar trajectory information is generated by the UAV motion parameters retrieved through the AirSim platform API. Accordingly, the structured information comprises the UAV identifiers (IDs), payload status, and confidence scores obtained from the visual modality, along with the position and velocity parameters of each UAV. After incorporating prior information via prompts, the MLLM evaluates the likely operational intent of adversarial UAVs and outputs reasoning results, as shown in Fig. \ref{Fig_3}.


\subsection{Structured Input}

The key to intent recognition lies in the ability to rapidly and accurately assess the state of adversarial UAVs. While UAVs can acquire RGB and infrared imagery and radar IQ data, transmitting such large-scale raw data directly imposes a substantial communication burden, particularly when multiple UAVs transmit simultaneously. Meanwhile, although MLLMs possess VFM and VLM components capable of understanding visual modalities, processing images requires computationally expensive encoding and cross-modal alignment, consuming significant resources. To address this, a structured data and key frame fusion strategy is proposed, in which onboard visual sensors, radar, and embedded visual classification models are utilized to extract UAV payload information. In contrast, radar provides trajectory and velocity data to form structured inputs for routine transmission. Only when the onboard visual classifier’s confidence score is low is the corresponding image sent to the cloud model for further analysis, thus reducing communication and computational overhead.

\subsection{Prompt-Assisted Mechanism}

A critical limitation of conventional intent recognition approaches is their dependence on fixed scenarios, with predefined label sets constraining the output space. Such methods lack flexibility in adapting to evolving battlefield conditions. In actual confrontations, situational dynamics are highly variable, and the same action may have different implications under different operational contexts. Compared with these methods, MLLMs offer substantial potential by dynamically adjusting tactical preferences according to the operational environment, enabling intent recognition across diverse mission types such as protection and interception. By inputting current scene descriptions, tactical preferences, and other relevant contextual data as prompts, the MLLM can incorporate prior knowledge to produce more accurate intent recognition results.

\subsection{Intent Recognition Results by MLLM}

Upon receiving the structured input, the MLLM analyzes the current UAV perception results. By considering both the motion dynamics and payload status of the two enemy UAVs, MLLM can infer their likely intentions, assess the potential threat level of each UAV, and predict their subsequent maneuvers. Importantly, beyond threat assessment, the MLLM generates multi-step action recommendations for the self side, providing valuable decision support for addressing the evolving situation. These recommendations can be further interpreted as potential control signals for the self UAV, thus bridging situational understanding with actionable guidance in a dynamic operational environment.

\section{Main Challenges and Future Work}

\subsubsection{High Dynamic Communication}

In aerial maneuvering scenarios, UAVs frequently adjust their positions to maintain optimal observation conditions. Under high-speed maneuvers, the channel environment changes rapidly, forming a typical high-dynamic communication scenario \cite{HD}. Frequent position shifts between transceivers also make coverage a critical issue, as limited coverage may cause data transmission failures and compromise the timeliness of intent recognition tasks. Thus, robust channel estimation algorithms and fast access strategies are essential. Leveraging UAVs’ real-time positional dynamics to switch between base stations or satellites can ensure more reliable coverage and stable data transmission.

\subsubsection{Multi-UAV Collaboration}

Global awareness can be achieved if the MLLM receives data from multiple UAVs simultaneously. Waiting for data transmitted from all self UAVs often introduces significant delays, causing intent recognition results to become outdated. Moreover, UAVs differ in observation perspectives, and data quality varies due to occlusion, distance, and sensor performance. Therefore, a dynamic data transmission and fusion strategy is needed to evaluate and rank UAV data in real time and transform multi-source observations into a unified reference frame with accurate association. Within limited transmission slots, priority should be given to UAVs with the highest-quality data, ensuring link capacity while enhancing the timeliness and reliability of intent recognition.

\subsubsection{Electromagnetic Spectrum Situational Awareness}

MLLM-enabled UAV intent recognition provides a new paradigm for low-altitude defense, but its robustness, due to the inaccuracy of the sensing data, still requires optimization. Meanwhile, electromagnetic spectrum situational awareness (EMSSA) offers broader coverage. However, EMSSA faces challenges such as complex modeling, spectrum congestion, and limited reliability under adversarial conditions. Future research should explore the integration framework of UAV intent recognition and spectrum situational awareness to achieve a more comprehensive, intelligent, and secure low-altitude defense system.

\subsubsection{Edge-Cloud Collaboration}

Achieving fast and accurate intent recognition remains a critical challenge. Assuming the entire task of intent recognition is offloaded to the cloud, the system must bear the high communication costs of transmitting large volumes of image data, and the alignment of multimodal data at the cloud side significantly increases inference delay. Therefore, edge-cloud collaboration is a promising direction for MLLM-enabled UAV intent recognition. Communication costs can be considerably reduced by effectively leveraging edge devices to summarize and abstract multimodal data and selectively transmit only critical information to the cloud. Simultaneously, the computational power of cloud infrastructure can be fully utilized to perform timely and accurate intent recognition.

\subsubsection{MLLM-Enabled UAV Scheduling}

MLLM-enabled UAV scheduling is expected to become a key research direction in low-altitude secure. The central objective is to leverage MLLMs to perform intent recognition and autonomously generate targeted countermeasure scheduling strategies, ultimately forming a unified {\it \textbf{perception-reasoning-scheduling }} closed-loop system. MLLM integrates the current threat assessment, operational rules, and the status of available countermeasure resources to generate concrete scheduling directives. These directives are then output as structured command streams, enabling the autonomous coordination of self UAVs to effectively respond to and neutralize adversarial UAV threats.

\section{Conclusion}

This paper presents an architecture for MLLM-enabled UAV intent recognition, comprising a multimodal fusion UAV perception module and an intent inference module based on MLLM. Then the basic knowledge of the architecture is introduced, and a review of the progress of research in multimodal UAV perception and UAV intent recognition is provided. Subsequently, a use case for low-altitude confrontation is presented, revealing the potential of applying MLLM to UAV intent recognition by demonstrating the intent recognition results achieved by MLLM. Finally, future challenges for MLLM-enabled UAV intent recognition are discussed, followed by corresponding strategic recommendations.

	\bibliographystyle{IEEEtran}
	\bibliography{reference}

\end{document}